\documentclass[sigconf]{acmart}

\AtBeginDocument{%
  \providecommand\BibTeX{{%
    \normalfont B\kern-0.5em{\scshape i\kern-0.25em b}\kern-0.8em\TeX}}}

\copyrightyear{2023} 
\acmYear{2023} 
\setcopyright{acmlicensed}\acmConference[GLSVLSI '23]{Proceedings of the Great Lakes Symposium on VLSI 2023}{June 5--7, 2023}{Knoxville, TN, USA}
\acmBooktitle{Proceedings of the Great Lakes Symposium on VLSI 2023 (GLSVLSI '23), June 5--7, 2023, Knoxville, TN, USA}
\acmPrice{15.00}
\acmDOI{10.1145/3583781.3590258}
\acmISBN{979-8-4007-0125-2/23/06}

\usepackage{multirow}
\usepackage[para]{threeparttable}

\begin{document}

\title{High-Speed and Energy-Efficient Non-Binary Computing with Polymorphic Electro-Optic Circuits and Architectures}


\author{Ishan Thakkar}
\affiliation{%
 \department{Department of ECE}
 \institution{University of Kentucky}
 \city{Lexington}
 \state{Kentucky}
 \country{USA}}
 \email{igthakkar@uky.edu}

\author{Sairam Sri Vatsavai}
\affiliation{%
 \department{Department of ECE}
 \institution{University of Kentucky}
 \city{Lexington}
 \state{Kentucky}
 \country{USA}}
\email{ssr226@uky.edu}

 \author{Venkata Sai Praneeth Karempudi}
\affiliation{%
 \department{Department of ECE}
 \institution{University of Kentucky}
 \city{Lexington}
 \state{Kentucky}
 \country{USA}}
\email{kvspraneeth@uky.edu}

\renewcommand{\shortauthors}{Ishan Thakkar, Sairam Sri Vatsavai, \& Venkata Sai Praneeth Karempudi}



\begin{abstract}
In this paper, we present microring resonator (MRR) based polymorphic E-O circuits and architectures that can be employed for high-speed and energy-efficient non-binary reconfigurable computing. Our polymorphic E-O circuits can be dynamically programmed to implement different logic and arithmetic functions at different times. They can provide compactness and polymorphism to consequently improve operand handling, reduce idle time, and increase amortization of area and static power overheads. When combined with flexible photodetectors with the innate ability to accumulate a high number of optical pulses in situ, our circuits can support energy-efficient processing of data in non-binary formats such as  stochastic/unary and high-dimensional reservoir formats. Furthermore, our polymorphic E-O circuits  enable configurable E-O computing accelerator architectures for processing binarized and integer quantized convolutional neural networks (CNNs). We compare our designed polymorphic E-O circuits and architectures to several circuits and architectures from prior works in terms of area, latency, and energy consumption. 

\end{abstract}






\maketitle

\section{Introduction}
In recent years, Moore's law has faced fatal challenges as the nanofabrication technology is experiencing serious limitations, due to the exceedingly small size of transistors \cite{amlan2022}. In the wake of dwindling Moore's law, fortunately, integrated electro-optic (E-O) computing circuits have shown the revolutionary potential to provide progressively faster and more efficient hardware for computing. The E-O circuits for computing, which have been demonstrated in prior works (e.g., \cite{lightbulb,pixel,shiflett2021,qiu2012,ying2019integrated,karempudi2021,ying2020}), are typically used to implement the following four types of logical and arithmetic functions: \textbf{(I)} \textbf{Basic logic-gate functions} \cite{lightbulb,shiflett2021,pixel} with two binary input operands that aid the acceleration of neural networks. \textbf{(II)} \textbf{Arbitrary combinational logic functions} \cite{qiu2012,ying2019integrated} that can work as the direct optical replacement of field programmable gate arrays (FPGAs). \textbf{(III)} \textbf{Two operand arithmetic functions} \cite{ying2020,karempudi2021} for accumulation that can support custom precision and full precision arithmetic operations. \textbf{(IV)} \textbf{Multi-operand linear arithmetic functions} \cite{lightbulb,holylight,deapcnn,pixel} to implement Multiply-Accumulate (MAC) and Vector Dot Product (VDP) operations for deep learning workloads. However, as elaborated in \cite{karempudi2023}, these E-O circuits face shortcomings due to their \textit{(i)} long idle time and resultant non-amortizable high area and static power overheads and \textit{(ii)} strong trade-off between wavelength parallelism and achievable bit-precision.

To alleviate these shortcomings, our contribution in this paper is two-fold: \textbf{(i)} Invention of a polymorphic E-O circuit (PEOC) and \textbf{(ii)} Design of a configurable E-O computing accelerator (CEONA). We show that our PEOC can be reconfigured to implement different arithmetic and logic functions at different times. Such PEOCs are employed in our CEONA in a wavelength division multiplexing (WDM) manner to provide flexible support for accelerating convolutional neural networks (CNNs) with various bit-precisions. Furthermore, CEONA enables the acceleration of delayed feedback reservoir computing (DFRC)-based applications. We compare our designed PEOC and CEONA to several circuits and architectures from prior works and show their benefits in terms of area, latency, and energy consumption. To gain preliminary knowledge before digging into the paper, we recommend the reader to go through the tutorials on microring resonators (MRRs) \cite{MRR-tutorial} and optical computing architectures \cite{deapcnn,SiPh-NN-AI-Survey}.

\begin{figure}[h!]
    \centering
    \includegraphics[width = \linewidth]{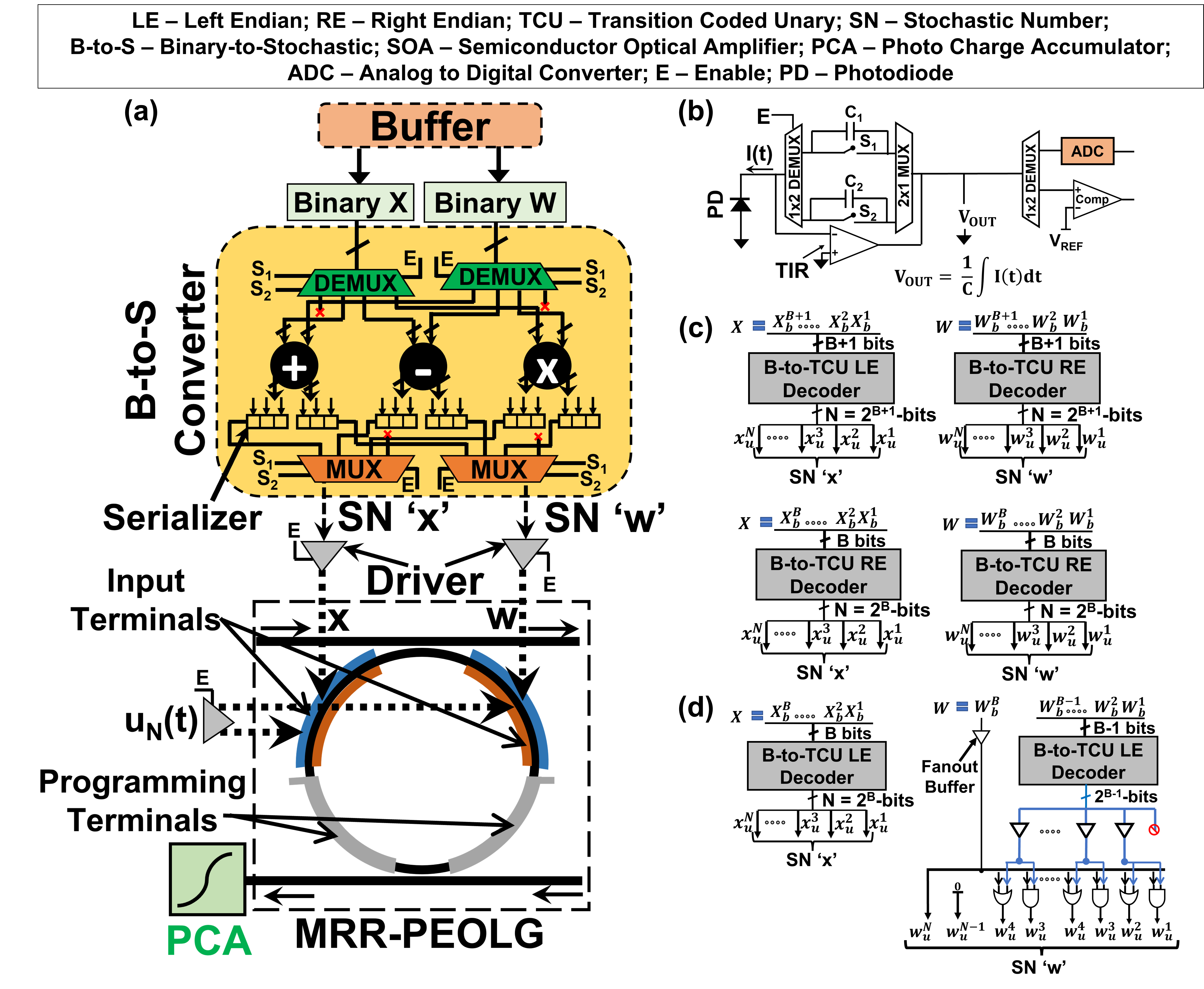}
    \caption{(a) Structure of our MRR-based polymorphic electro-optic circuit (PEOC), (b) photo-charge accumulator (PCA). B-to-S conversion circuits for ADD ((c)-top), SUB ((c)-bottom), and MUL (d) operations.}
    \label{unary}
\end{figure}

\vspace{-5pt}

\section{Polymorphic Electro-Optic Circuit} \label{sec2}
Fig. \ref{unary}(a) illustrates the structure of our polymorphic electro-optic (E-O) circuit (PEOC). It consists of an active MRR that can be utilized as either a microring modulator (MRM) or a polymorphic E-O logic gate (MRR-PEOLG) \cite{karempudi2023}. Using the active MRR as an MRM enables modulation of an incoming, electrical time-series signal u$_N$(t) onto an output optical signal \cite{SairamVLSID2021}. Moreover, when the active MRR is used as an MRR-PEOLG, it can be reconfigured to implement different logic functions at different times \cite{karempudi2023}. All of the AND, OR, XOR, NAND, NOR, and XNOR logic functions have been demonstrated \cite{karempudi2023}. In addition, in Fig. \ref{unary}, a binary-to-stochastic (B-to-S) conversion circuit and a photo-charge accumulator (PCA) are integrated with the MRR-PEOLG to transform the PEOC into a polymorphic binary arithmetic unit (PBAU). PBAU leverages the OR, AND, and XOR functions of the MRR-PEOLG to implement stochastic computing (SC) based approximate addition (ADD), multiplication (MUL), and subtraction (SUB) operations \cite{alaghi2013}. 
More details on the structure and operation of our MRR-PEOLG, PCA, and PBAU are provided in the upcoming subsections.



\begin{figure*}[h!]
    \centering
    \includegraphics[width = \linewidth]{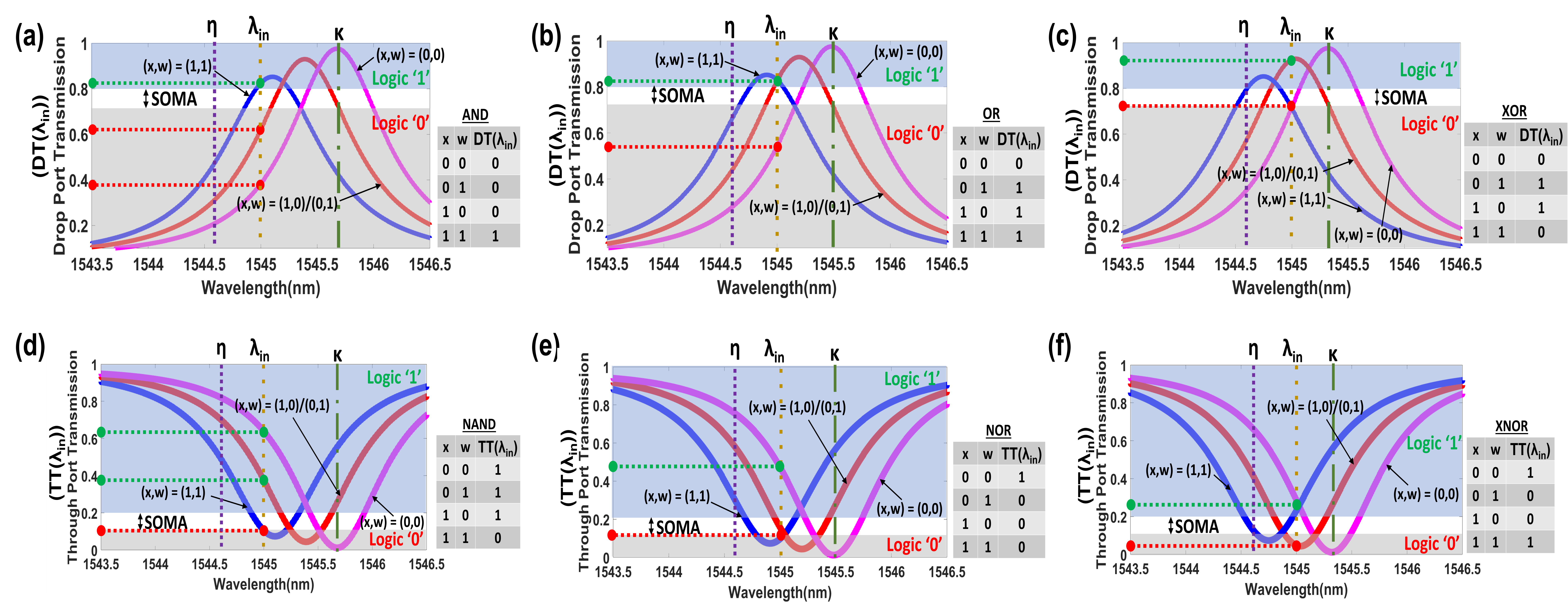}
    \caption{Transmission spectra of our MRR-PEOLG from \cite{karempudi2023} for (a) AND, (b) OR, (c) XOR, (d) NAND, (e) NOR, (f) XNOR.}
    \label{Freq}
\end{figure*}

\begin{table}[h!]
\centering
\caption{Performance comparison of E-O circuits.}
\begin{tabular}{|c|cc|cc|}
\hline
\multirow{2}{*}{\textbf{Metrics}} & \multicolumn{2}{c|}{\textbf{XNOR-POPCOUNT}}             & \multicolumn{2}{c|}{\textbf{Bit-serial Multiplier}}         \\ \cline{2-5} 
                         & \multicolumn{1}{c|}{\cite{lightbulb}} & \textbf{MRR-PEOLG}           & \multicolumn{1}{c|}{\cite{shiflett2021}} & \textbf{MRR-PEOLG}            \\ \hline
A (mm$^2$)                  & \multicolumn{1}{c|}{0.013}    & 0.011 (1.16×)  & \multicolumn{1}{c|}{0.023}    & 0.011 (2.08×)   \\ \hline
E (nJ)                   & \multicolumn{1}{c|}{0.05}     & 0.032 (1.53×)  & \multicolumn{1}{c|}{0.327}    & 0.033 (9.89×)   \\ \hline
L (ns)                   & \multicolumn{1}{c|}{0.02}     & 0.025 (0.8×)   & \multicolumn{1}{c|}{0.1}      & 0.025 (4×)      \\ \hline
A*E*L                    & \multicolumn{1}{c|}{1.3e-5}   & 0.9e-5 (1.44×) & \multicolumn{1}{c|}{75.2e-5}  & 0.91e-5 (82.6×) \\ \hline
\end{tabular}
\label{table2}
\end{table}

\subsection{MRR-Based Polymorphic E-O Logic Gate} \label{sec2_1}
\begin{figure}[h!]
    \centering
    \includegraphics[width = \linewidth]{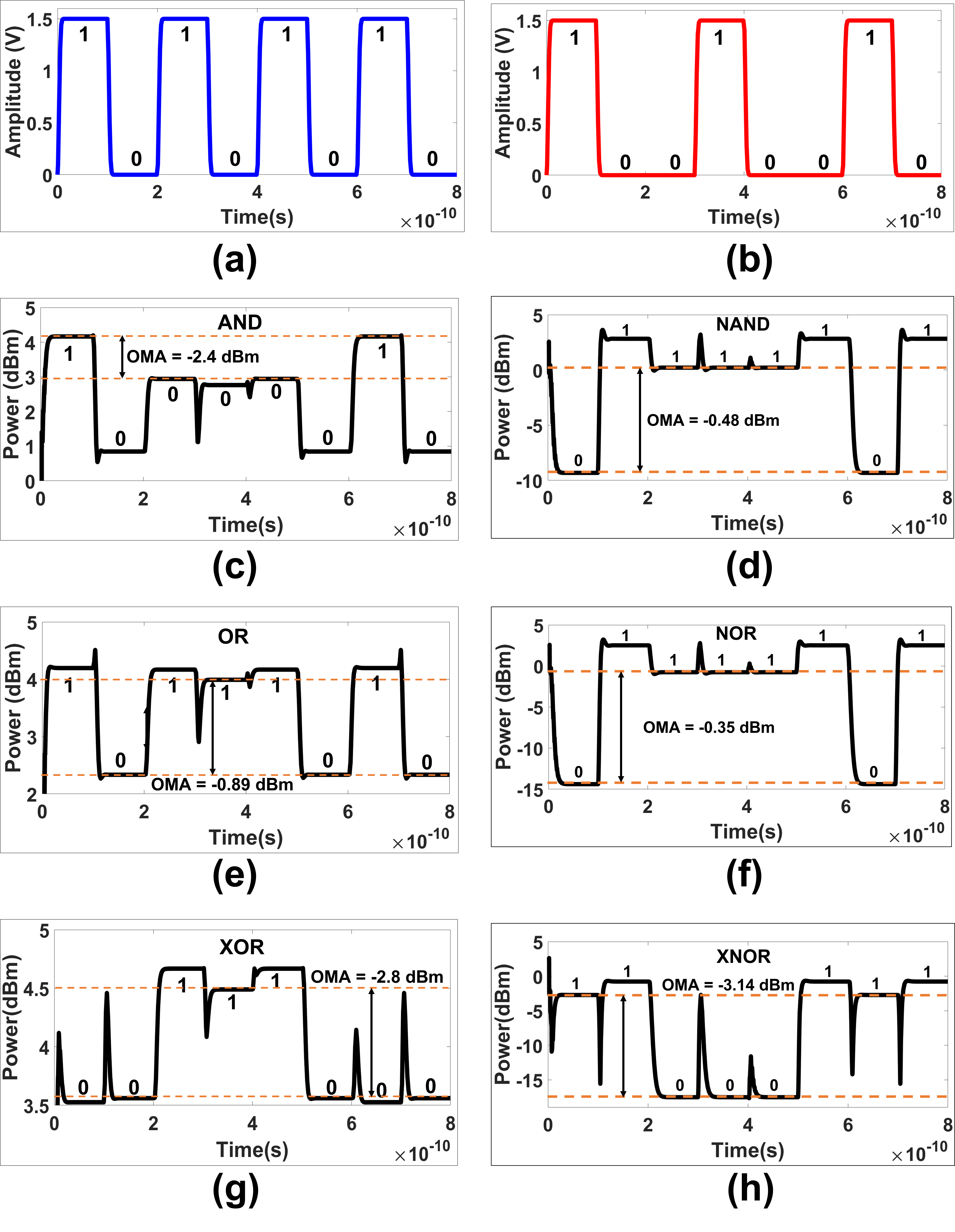}
    \caption{Transient analysis results for our MRR-PEOLG from \cite{karempudi2023}. (a),(b) Input electrical pulses. Output optical pulses for (c) AND, (d) NAND, (e) OR, (f) NOR, (g) XOR, (h) XNOR.}
    \label{time}
\end{figure}

Our invented MRR-PEOLG is described in \cite{karempudi2023}. From \cite{karempudi2023}, to program MRR-PEOLG to implement a specific logic-gate function, the MRR's operand-independent resonance position '$\kappa$' (magenta-colored passband in Fig. \ref{Freq}) is adjusted to a specific spectral position with respect to the input wavelength '$\lambda$$_{in}$' and the MRR's initial resonance position '$\eta$', by applying a voltage to the programming terminals of the MRR-PEOLG (see the terminals in Fig. \ref{unary}(a)). Then, the electrical input operands are applied to the PN junction-based input terminals of the MRR ((x,w) in Figs. \ref{unary} and \ref{Freq}). Upon doing so, the resonance of the MRR shifts towards shorter wavelengths depending on the combination of applied input operands. Applying the input operand bits to the input terminals makes the drop-port and through-port optical responses of our MRR-PEOLG follow the truth table of logic gate functions for which the MRR-PEOLG is programmed. In this manner, our MRR-PEOLG can perform different logic functions at different times (Fig. \ref{Freq}). To validate this polymorphic functionality of our MRR-PEOLG, we also performed a time-domain (transient) analysis using the INTERCONNECT tool of Ansys/Lumerical suite \cite{lumerical}. For that, we provided two electrical pulses (Figs. \ref{time}(a) and \ref{time}(b)) to the input terminals of our MRR-PEOLG and collected the output pulse patterns corresponding to different logic functions at the drop-port (Figs. \ref{time}(c), \ref{time}(e) and \ref{time}(g)) and through-port (Figs. \ref{time}(d), \ref{time}(f) and \ref{time}(h)) of our MRR-PEOLG. As evident, the output signals follow the pulse-wise truth tables of the respective logic functions, which demonstrates the capability of our MRR-PEOLG to implement different logic functions. In addition, we evaluated how the use of our MRR-PEOLG improves the area, latency, and energy consumption of two E-O circuits from prior works \cite{lightbulb} and \cite{shiflett2021}. The results are provided in Table \ref{table2}. 


\subsection{Photo-Charge Accumulator}\label{sec2_3}
Fig. \ref{unary}(b) illustrates our Photo-Charge Accumulator (PCA), which is  collectively inspired by the time integrating receiver (TIR) from \cite{alexandermit2022} and the photodetector (PD)-based optical-pulse accumulator from \cite{BhaskaranPCA2022}. Hence, our PCA employs a PD, a TIR, and two capacitors. The ADC or comparator is connected to the TIR based on the intended use case. 
The PD has the ability of dual coherent-incoherent superposition and photo-charge accumulation \cite{BhaskaranPCA2022}. If the PD has inverse bandwidth of \textit{t}=(1/symbol rate), the number of free carriers generated in the PD during every interval of \textit{t} is proportional to the number of photons absorbed, and hence, the output photocurrent in the interval is proportional to the sum of average optical powers of all the coherent and incoherent optical pulses that are incident on the PD during the interval \cite{BhaskaranPCA2022}. This output photocurrent is collected by the TIR of our PCA to generate a proportional voltage on one of the capacitors. This output photocurrent, and hence the voltage accrued on the capacitor, provides the accumulation result of all the optical pulses that are incident on the PD during the interval of time \textit{t} \cite{alexandermit2022}. The TIR circuit can allow the accumulation of a total of $\gamma$ such intervals of time \textit{t} each before the PCA circuit saturates. $\gamma$ is known as PCA's accumulation capacity. After the optical pulse accumulation over $\gamma$ intervals, a  discharge of the active capacitor (e.g., C1) is needed to prepare the circuit for the next accumulation. While capacitor C1 is discharging, capacitor C2 mitigates the discharge latency by allowing a continuation of another concurrent accumulation. Table \ref{pcascalability} reports our PCA's accumulation capacity $\gamma$ at different symbol rates from \cite{oxbnn}. At 50 GS/s, our PCA has $\gamma$=8503 (Table \ref{pcascalability}), which is greater than the required accumulation count per neuron for most modern CNNs. Such large $\gamma$ for our PCA eliminates the need to decompose the required accumulations per output neuron into multiple partial sums \cite{oxbnn}.

\vspace{5pt}

\begin{table}[h!]
\caption{PCA accumulation capacity ($\gamma$) for different symbol rates (SRs)  (GS/s) \cite{oxbnn}.}
\label{pcascalability}
\begin{tabular}{|c|c|c|c|l|c|l|c|}
\hline
\textbf{SR }    & 3     & 5     & 10    & 20    & 30    & 40   & 50   \\ \hline
\textbf{$\gamma$} & 39682 & 29761 & 19841 & 14880 & 10822 & 9920 & 8503 \\ \hline
\end{tabular}
\label{table4}
\end{table}

\subsection{Polymorphic Binary Arithmetic Unit}
From Section \ref{sec2} and Fig. \ref{unary}, integrating our MRR-PEOLG with a B-to-S conversion circuit and a PCA transforms the PEOC into a polymorphic binary arithmetic unit (PBAU). The structure of our PBAU can be segmented into three stages. First, the B-to-S peripheral circuit aids in converting the input binary operands (N-bit) into stochastic bit-streams (2$^{N}$-bits). The B-to-S conversion, in a nutshell, is implemented through bit-parallel binary-to-transition coded unary (B-to-TCU) decoders \cite{sri2023bit}. The bit-parallel outputs of B-to-TCU decoders are then converted into bit-streams using high-speed serializers. The B-to-TCU decoders are custom designed for ADD, MUL, and SUB functions, to ensure that the input stochastic bit-streams ('x' and 'w') have an appropriate correlation to minimize the errors in their results \cite{alaghi2013}. To achieve appropriate correlation among the stochastic bit-streams, we learned from \cite{alaghi2013} to endow the function-specific B-to-S circuits (and the constituent B-to-TCU decoders) with appropriate endianness and bit-stream sizes. For instance, for ADD functions, bit-streams x and w have opposite endianness, whereas for SUB and MUL they both have the same endianness (right endianness) (Figs. \ref{unary}(c) and \ref{unary}(d)). Moreover, the generated stochastic bit-streams from the B-to-S conversion circuits have 2$^{N}$ bits for MUL and SUB functions, whereas they have 2$^{N+1}$ bit for ADD function \cite{alaghi2013}. Further, our B-to-S conversion circuit for MUL function (Fig. \ref{unary}(d)) minimizes correlation-related errors in the results by ensuring that the conditional probability P(w/x) is equal to the marginal probability P(x) \cite{sri2023bit}.

The stochastic bit-streams generated from the B-to-S conversion stage are given as input to the second stage, which consists of our MRR-PEOLG \cite{karempudi2023}. Our MRR-PEOLG as described in Section \ref{sec2_1} implements AND, OR, and XOR logical functions, which are applied to the stochastic bit-streams in a bit-wise manner to implement the target MUL, ADD, and SUB functions respectively. 
The third stage of PBAU consists of our PCA, which converts the resultant stochastic bit-stream from the MRR-PEOLG into the binary format.

\vspace{3pt}

\begin{table}[h!]
\caption{Performance of our PBAU for ADD, SUB, and MUL. \\MAE=Mean Absolute Error.}
\begin{tabular}{|c|ccc|ccc|}
\hline
\multirow{2}{*}{Bit Precision}                                       & \multicolumn{3}{c|}{6-bit}                                   & \multicolumn{3}{c|}{8-bit}                                      \\ \cline{2-7} 
                                                                     & \multicolumn{1}{c|}{ADD}  & \multicolumn{1}{c|}{SUB}  & MUL  & \multicolumn{1}{c|}{ADD}   & \multicolumn{1}{c|}{SUB}   & MUL   \\ \hline
Latency (ns)                                                         & \multicolumn{1}{c|}{5.32} & \multicolumn{1}{c|}{2.74} & 2.76 & \multicolumn{1}{c|}{20.51} & \multicolumn{1}{c|}{10.27} & 10.29 \\ \hline
Energy (pJ)                                                          & \multicolumn{1}{c|}{16.1} & \multicolumn{1}{c|}{6.8}  & 10.2 & \multicolumn{1}{c|}{60.1}  & \multicolumn{1}{c|}{23.6}  & 36.2  \\ \hline
\begin{tabular}[c]{@{}c@{}}MAE\end{tabular} & \multicolumn{1}{c|}{0}    & \multicolumn{1}{c|}{0}    & 0.03 & \multicolumn{1}{c|}{0}     & \multicolumn{1}{c|}{0}     & 0.04  \\ \hline
\end{tabular}
\label{table3}
\end{table}

\vspace{3pt}

\begin{table}[h!]
\centering
\caption{Comparison of latency, energy and area of PBAU with E-O arithmetic circuits from prior work.}
\begin{tabular}{|c|c|c|c|}
\hline
 &Area & Energy       & Area*Latency          \\ \hline
8-bit PBAU      & 0.0012mm$^2$  &36.2pJ      &3.312mm$^2$.ps \\ \hline
8-bit PoNALU \cite{karempudi2021}       & 0.6 mm$^2$  &31.25nJ &201.3mm$^2$.ps   \\ \hline
8-bit EPALU \cite{ying2020}       &1.4 mm$^2$   &37.5nJ &523.5 mm$^2$.ps  \\ \hline
8-bit PIXEL \cite{pixel}    & 0.00359 mm$^2$   & 51.2pJ &36.9mm$^2$.ps  \\ \hline
\end{tabular}
\label{table4}
\end{table}

\vspace{3pt}

In Table \ref{table3}, we have evaluated the per-operation latency, energy, and mean absolute error (MAE) values of our PBAU for MUL, SUB, and ADD functions across the binary (integer) operand precision of 6-bit and 8-bit. As evident, our PBAU incurs no errors for SUB and ADD functions, and the MAE values for MUL function are also negligibly low. 
Similarly, Table \ref{table4} provides a comparison of latency, energy, and area of our PBAU with E-O arithmetic circuits from prior works. From Table \ref{table4}, our PBAU consumes substantially less energy and occupies less area.
\vspace{7pt}

\begin{figure}[h!]
    \centering
    \includegraphics[width=\linewidth]{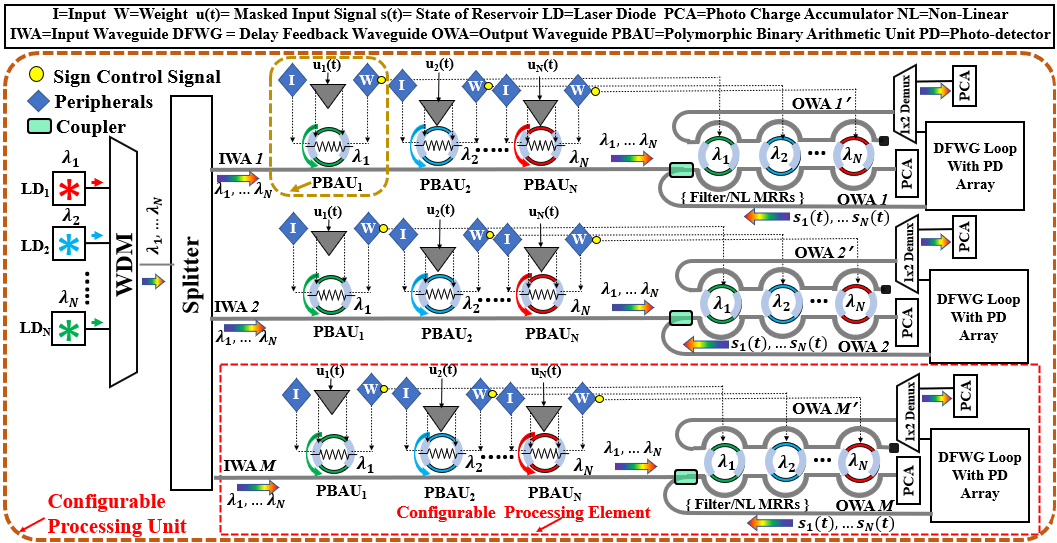}
    \caption{Schematic of our CEONA architecture.}
    \label{CEONA}
\end{figure}

\section{Configurable E-O Computing Accelerator}
\subsection{Overview}
Multiple PBAUs are organized in an array employing wavelength division multiplexing (WDM) to constitute the main processing unit of our Configurable E-O computing Accelerator (CEONA) architecture. This unit is called a configurable processing unit (CoPU), as illustrated in Fig. \ref{CEONA}. A CoPU consists of a comb laser source \cite{praneeth4pam} that emits optical power at \textit{N} distinct wavelengths (i.e., from $\lambda_1$ to $\lambda_N$). The optical power at each of these wavelengths is split into \textit{M} input waveguides (IWA), each of which is connected to a configurable processing element (CoPE). Each CoPE consists of an array of \textit{N} PBAUs arranged in a WDM manner aside a bank of MRRs that act as filters or non-linear active MRRs (Fig. \ref{CEONA}). Each  MRR bank connects to PCAs or a delay feedback loop waveguide (DFWG). Based on the configuration of the constituent PBAUs and MRR banks, and CoPE's connection to PCAs or DFWG, our CEONA accelerator can be employed in two use cases.

\subsection{Case I: Neural Network Accelerator}
CEONA can be configured to perform inference of binary neural networks (BNNs) and integer-quantized CNNs. During inference, CEONA receives weight and input operands that are 1-bit (binarized) for BNNs and 8-bit for integer-quantized CNNs.  

\textbf{CEONA with Binarized Operands:} CEONA with binarized operands is referred to as CEONA-B. The inference of BNNs requires XNOR-Bitcount operations \cite{oxbnn}. Therefore, with binarized operands, CEONA dynamically configures each CoPE's PBAUs as XNOR gates as discussed in Section \ref{sec2}. The PBAUs perform the XNOR operation between binarized I and W (Fig. \ref{CEONA}). The optical outputs of XNOR gates are sent to the bottom OWA (Fig. \ref{CEONA}). The MRR banks are turned off to allow all XNOR output bits to reach the bottom PCA (on OWA). The PCA counts the incoming optical bits coming from PBAUs to do bitcount operations to generate final accumulation results. Thus, each CoPE can generate one value of the output BNN tensor without requiring partial sum storage or reduction, as detailed in \cite{oxbnn}. Our utilized evaluation setup is reported in \cite{oxbnn}.       

\begin{figure}[h!]
  \centering
  \includegraphics[width=\linewidth]{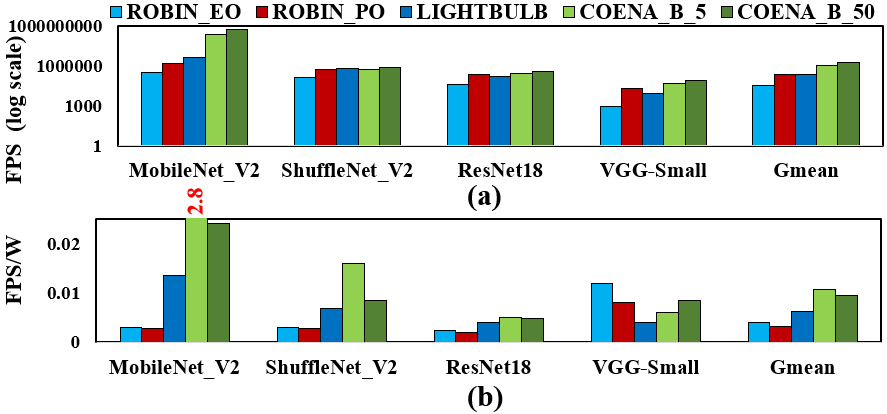}
  \caption{(a) FPS (log scale), (b) FPS/W for CEONA-B versus ROBIN \cite{robin} and LIGHTBULB \cite{lightbulb} accelerators.} 
  \label{CEONA-B}
\end{figure}

Figs. \ref{CEONA-B}(a) and \ref{CEONA-B}(b) compare FPS (Frames-Per-Second)(throughput) values and FPS/W (energy-efficiency) values achieved by CEONA-B and prior accelerators across various BNNs. Overall, both CEONA-B\_5 (SR=5 GS/s) and CEONA-B\_50 (SR=50 GS/s) achieve better throughput and energy efficiency than other accelerators. CEONA-B\_50 achieves 52$\times$, 7$\times$, and 7$\times$ better FPS than ROBIN\_EO \cite{robin}, ROBIN\_PO \cite{robin}, and LIGHTBULB \cite{lightbulb}, respectively, on gmean across the BNNs. Our CEONA-B\_5 gains 2.6$\times$, 3.3$\times$, and 1.7$\times$ better FPS/W than ROBIN\_EO, ROBIN\_PO, and LIGHTBULB, respectively, on gmean across the BNNs. From \cite{oxbnn}, CEONA-B improves throughput by achieving larger \textit{N} resulting in higher parallelism. The energy benefits come mainly from the elimination of the need to store and reduce partial sums due to the PCAs' innate capability of performing in-situ accumulations. 

\vspace{5pt}

\textbf{CEONA with Integer Operands:} CEONA with integer operands is referred to as CEONA-I. The inference of CNNs requires dot product operations \cite{cases2022}. As discussed in Section \ref{sec2}, with stochastic computing, multiplication can be performed with AND gates. Therefore, for inference of CNNs, CEONA-I configures PBAUs to work as AND gates and they perform pointwise multiplication of I and W \cite{sconna} (Fig. \ref{CEONA}). The MRR banks are operated as filter banks, and sign control signals from corresponding PBAUs turn on/off the filters to enable signed accumulation at the PCAs. A comprehensive explanation of the CEONA-I architecture and employed evaluation setup is provided in \cite{sconna}.

\begin{figure}[h!]
  \centering
  \includegraphics[width=\linewidth]{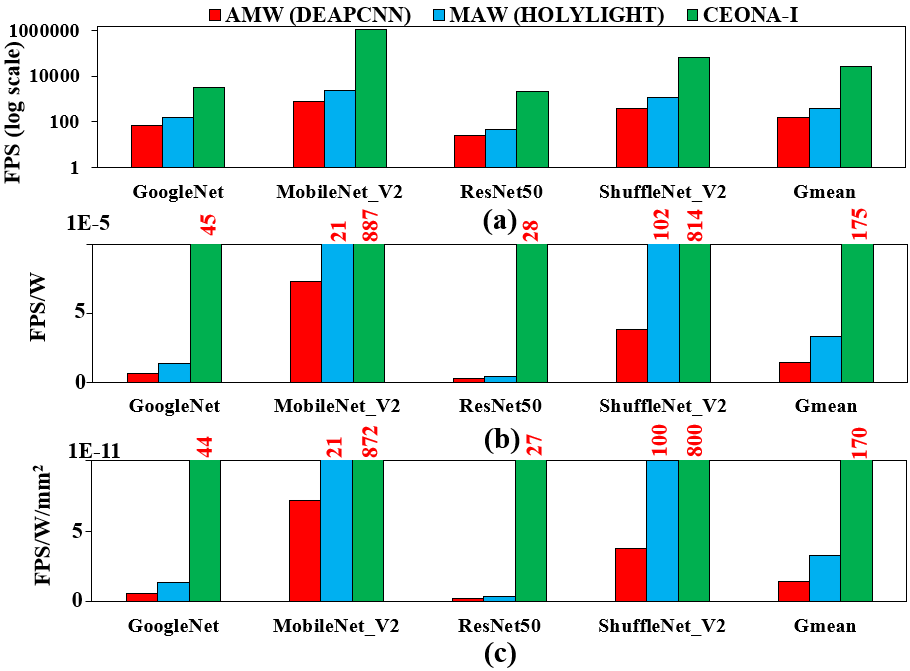}
  \caption{(a) FPS, (b) FPS/W, (c) FPS/W/$mm^2$ for CEONA-I versus MAW and AMW accelerators for 8-bit integer precision.} 
  \label{CEONA_I}
\end{figure}

Figs. \ref{CEONA_I}(a), \ref{CEONA_I}(b), and \ref{CEONA_I}(c) compare the FPS (throughput) values, FPS/W (energy efficiency), and FPS/W/mm$^2$ (area efficiency) achieved by CEONA-I and prior accelerators across various CNNs. From Fig. \ref{CEONA_I}(a),  CEONA-I significantly outperforms the analog optical accelerators MAW (HOLYLIGHT) \cite{holylight} and AMW (DEAPCNN) \cite{deapcnn} by 66.5$\times$ and 146.4$\times$, respectively, on gmean across the CNNs.  From Fig. \ref{CEONA_I}(b), CEONA-I gains 90$\times$ and 183$\times$ better FPS/W than analog MAW (HOLYLIGHT) and AMW (DEAPCNN), respectively, on gmean across the CNNs. From Fig. \ref{CEONA_I}(c), CEONA-I gains 91$\times$ and 184$\times$ better FPS/W/mm$^2$ than analog MAW (HOLYLIGHT) and AMW (DEAPCNN), respectively, on gmean across the CNNs. The throughput benefits are mainly associated with the superior \textit{N} of CEONA-I compared to the analog optical accelerators \cite{sconna}. Moreover, the use of PCAs eliminates the need to store and reduce partial sums, providing throughput, energy, and area benefits. Overall, CEONA-I significantly improves throughput, energy efficiency, and area efficiency compared to prior analog accelerators.  

\vspace{5pt}

\begin{figure}[h!]
  \centering
  \includegraphics[scale=0.39]{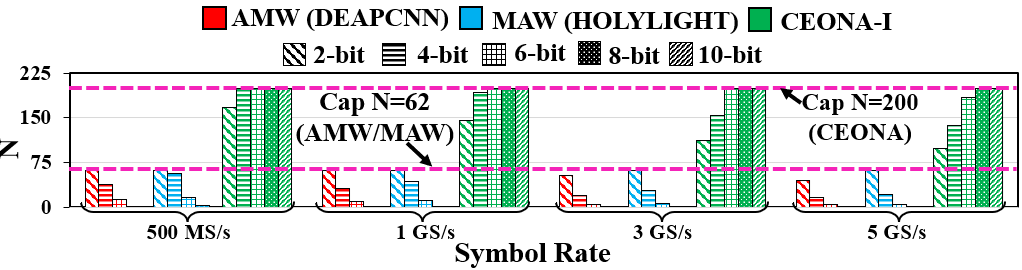}
  \caption{Supported CoPE size \textit{N} for bit precision =\{2, 4, 6, 8, 10\}bits at symbol rates (SRs) = \{0.5, 1, 3, 5\}GS/s, for AMW, MAW, and  CEONA-I.} 
  \label{CEONA_I_Scalability}
\end{figure}

\textbf{Scalability of CEONA-I:} To determine the achievable size \textit{N} for our CEONA-I CoPU across various integer precision levels (B), we adopt scalability analysis equations (Eq. \ref{eq1}, Eq. \ref{eq2}, and Eq. \ref{eq3}) from \cite{lukasscalability} and \cite{cases2022}. The definitions of the parameters and their values used in these equations are reported in \cite{sconna}. In Eq. \ref{eq1}, for AMW and MAW architectures, the datarate (DR) is equal to the symbol rate (SR) and $n_{i/p}$=B. But, for CEONA-I architecture, DR=(SR/$2^{B}$) and $n_{i/p}$=1 \cite{sconna}. We consider \textit{M=N} and solve the equations using the method from \cite{sconna}. Fig. \ref{CEONA_I_Scalability} reports the achievable \textit{N} of CEONA-I, AMW, and MAW architectures for different B levels across various \textit{SR}s. As evident from Fig. \ref{CEONA_I_Scalability}, our CEONA-I can support larger \textit{N} value compared to AMW and MAW at all bit-precision levels across different SRs. For instance, CEONA-I achieves larger \textit{N=192} for 4-bit precision at 1 GS/s, compared to AMW and MAW, which achieve \textit{N=31} and \textit{N=44}, respectively. This is because of CEONA-I's PCAs' in-situ accumulation capacity \cite{BhaskaranPCA2022} (see Section \ref{sec2_3}). It allows the PCAs to operate at lower DR=(SR/2$^B$) which significantly improves \textit{N} at larger \textit{$B$}. In contrast, the support for \textit{N} decreases for AMW and MAW with an increase in $B$ \cite{cases2022}. Furthermore, the achievable \textit{N} is also limited by inter-wavelength spacing. We consider optimistic FSR=50nm. For AMW and MAW, the inter-wavelength spacing is set to 0.8nm \cite{cases2022} whereas for CEONA-I it can be set to 0.25nm \cite{sconna}. Therefore, in Fig. \ref{CEONA_I_Scalability}, the \textit{N} values for AMW/MAW and CEONA-I are capped at 62 (=FSR/0.8nm) and 200 (=FSR/0.25nm) respectively.   

\vspace{5pt}

\begin{equation}
      n_{i/p} = \frac{1}{6.02}\Bigg[20log_{10}(\frac{R\times P_{PD-opt}}{\beta\sqrt{\frac{DR}{\sqrt{2}}}}-1.76\Bigg]
      \label{eq1}
\end{equation}

\begin{equation}
    \beta = \sqrt{2q(RP_{PD-opt}+I_d)+\frac{4kT}{R_L}+R^2P_{PD-opt}^2RIN}
    \label{eq2}
\end{equation}

\begin{equation}
  \label{eq3}
\begin{split}
 P_{Laser} = \frac{10^{\frac{\eta_{WG}(dB)[N(d_{OSM})]}{10}}M}{\eta_{SMF}\eta_{EC}IL_{i/p-OSM}}
    \times\frac{P_{PD-opt}}{\eta_{WPE}IL_{MRR}}
      \\\times\frac{1}{(OBL_{OSM})^{N-1}(EL_{splitter})^{log_{2}M}} 
    \\ \times \frac{1}{(OBL_{MRR})^{N-1}(IL_{penalty})}
\end{split}
\end{equation}
\subsection{Case II: Reservoir Computing}
CEONA can also be configured as a delay feedback reservoir computing accelerator (CEONA-DFRC) for training and inference of time series tasks. For that, CEONA-DFRC configures the PBAUs to work as conventional modulators to modulate input masked signals \textit{u$_i$(t)} (Fig. \ref{CEONA}). Each MRR in the filter banks is configured to act as a non-linear node of the reservoir \cite{Appeltant2011InformationPU}. 

An active MRR shows the rich non-linear response at its drop-port transmission due to Two-Photon Absorption (TPA) \cite{BahadoriDATE2017}. The degree of non-linearity  depends on the photon lifetime ($\tau_{ph}$) of the MRR cavity. For an MRR, $\tau_{ph}$ depends on the MRR’s Q-factor. Therefore, the non-linearity of the MRR can be controlled with the Q-factor (hence, $\tau_{ph}$) of the MRR. To enable control of the MRRs' Q-factor (hence, $\tau_{ph}$), we employ the non-linear MRR design from \cite{ShomanOFC2018,Luan2023Nature} in Fig. \ref{CEONA}. 

In Fig. \ref{CEONA}, the non-linear MRRs along with the DFWG loop form a reservoir \cite{SairamVLSID2021}.  
The MRRs' quality factor is adapted \cite{Luan2023Nature,proteusnocs2020} to set the degree of nonlinearity depending on the task. The MRRs generate the states of the reservoir by a non-linear transformation of modulated u$_i(t)$. The states s$_i(t)$ of the reservoir are sent to the DFWG loop where they are captured and stored using the PD array, and part of s$_i(t)$ signals are further fed as feedback to the non-linear MRRs with modulated masked inputs u$_i(t+1)$. The operation and architecture of CEONA-DFRC are discussed extensively in \cite{SairamVLSID2021}.

\begin{figure}[h!]
  \centering
  \includegraphics[scale=0.42]{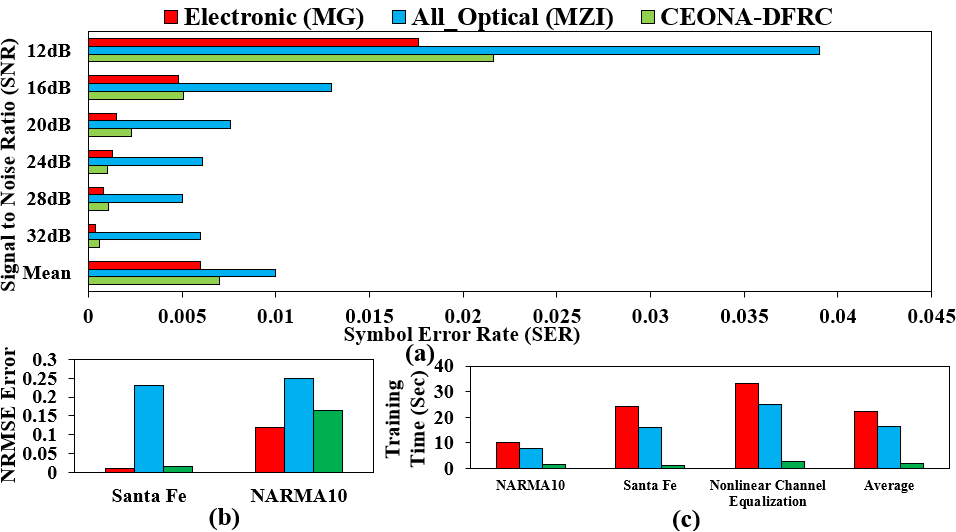}
  \caption{Performance evaluation and comparison of CEONA-DFRC, reproduced from \cite{SairamVLSID2021}. (a) SER for various target SNRs for channel equalization task, (b) NEMSE for Santa Fe and NARMA series prediction, and (c) training time comparison for considered tasks.} 
  \label{CEONA_DFRC}
\end{figure}

Figs. \ref{CEONA_DFRC}(a), \ref{CEONA_DFRC}(b), and \ref{CEONA_DFRC}(c) compare Symbol Error Rate (SER), Normalized Root Mean Square Error (NRMSE), and training time of various DFRC accelerators across various time series tasks. From Fig. \ref{CEONA_DFRC}(a), on average across various target SNRs, CEONA-DFRC achieves 58.8\% lower SER than All\_Optical (MZI) \cite{opticalmziDFRC} on non-linear channel equalization task \cite{channelequalization}. Similarly, Fig. \ref{CEONA_DFRC}(b) shows that for NARMA \cite{narma10} and SantaFe \cite{santfe}, CEONA-DFRC achieves 35\% lower NRMSE compared to All\_Optical (MZI), and it performs on par with Electronic (MG) \cite{Appeltant2011InformationPU}. CEONA-DFRC's major benefit can be observed in Fig. \ref{CEONA_DFRC}(c); it significantly speedups training time by 98$\times$ and 93$\times$ on average compared to All\_Optical (MZI) and Electronic (MG) respectively. CEONA-DFRC leverages the rich non-linearity of the active MRR to realize the non-linear node in the reservoir layer. Moreover, the MRR-based reservoir for CEONA-DFRC takes 168$\times$ and 2*$10^5$$\times$ less time to transform the masked input signal compared to All\_Optical MZI and Electronic MG reservoirs.  respectively. In addition, CEONA-DFRC uses a photonic waveguide as the delay feedback loop which further reduces the training time \cite{SairamVLSID2021}. Overall, CEONA-DFRC significantly improves the training time of time series tasks while achieving better or on-par error compared to prior optical and electronic DFRCs.

\vspace{5pt}

\vspace{-5pt}

\section{Summary and Future Prospects}
In this paper, we demonstrated our invented PEOC and CEONA. We showed that PEOC can be reconfigured to implement different logical and arithmetic functions at different times. Similarly, we showed that CEONA can be used to enable flexible support for accelerating CNNs and DFRC applications. We have shown that our CEONA can be reconfigured to handle binarized CNNs and integer-quantized CNNs. In both cases, CEONA provides significant benefits in throughput, energy efficiency, and area efficiency compared to prior works. In addition, we have also shown that CEONA achieves significant latency benefits for accelerating inference and training of various time series tasks such as NARMA10, SantaFe, and Non-linear Channel Equalization. 

We envision that our CEONA can be extended to support the acceleration of mixed-precision CNNs \cite{Chen_2021_ICCV}. Furthermore, each CoPE of our CEONA architecture can accelerate multiple time series tasks in a WDM manner to significantly improve the accelerator throughput. Overall, this flexibility can enable CEONA to simultaneously handle multiple workloads from machine learning and artificial intelligence applications.

\begin{acks}
This research is supported by a grant from NSF (CNS-2139167).
\end{acks}

\bibliographystyle{ACM-Reference-Format}
\bibliography{ref}


\begin{thebibliography}{35}


\ifx \showCODEN    \undefined \def \showCODEN     #1{\unskip}     \fi
\ifx \showDOI      \undefined \def \showDOI       #1{#1}\fi
\ifx \showISBNx    \undefined \def \showISBNx     #1{\unskip}     \fi
\ifx \showISBNxiii \undefined \def \showISBNxiii  #1{\unskip}     \fi
\ifx \showISSN     \undefined \def \showISSN      #1{\unskip}     \fi
\ifx \showLCCN     \undefined \def \showLCCN      #1{\unskip}     \fi
\ifx \shownote     \undefined \def \shownote      #1{#1}          \fi
\ifx \showarticletitle \undefined \def \showarticletitle #1{#1}   \fi
\ifx \showURL      \undefined \def \showURL       {\relax}        \fi
\providecommand\bibfield[2]{#2}
\providecommand\bibinfo[2]{#2}
\providecommand\natexlab[1]{#1}
\providecommand\showeprint[2][]{arXiv:#2}

\bibitem[A.{Ganguly \em et al.}(2022)]%
        {amlan2022}
\bibfield{author}{\bibinfo{person}{A.{Ganguly \em et al.}}}
  \bibinfo{year}{2022}\natexlab{}.
\newblock \showarticletitle{Interconnects for DNA, Quantum, In-Memory, and
  Optical Computing: Insights From a Panel Discussion}.
\newblock \bibinfo{journal}{\emph{IEEE micro}} \bibinfo{volume}{42},
  \bibinfo{number}{3} (\bibinfo{year}{2022}), \bibinfo{pages}{40--49}.
\newblock


\bibitem[Al-Qadasi et~al\mbox{.}(2022)]%
        {lukasscalability}
\bibfield{author}{\bibinfo{person}{M.~A. Al-Qadasi} {et~al\mbox{.}}}
  \bibinfo{year}{2022}\natexlab{}.
\newblock \showarticletitle{Scaling up silicon photonic based accelerators:
  Challenges and opportunities}.
\newblock \bibinfo{journal}{\emph{APL Photonics}} (\bibinfo{year}{2022}).
\newblock


\bibitem[Alaghi and Hayes(2013)]%
        {alaghi2013}
\bibfield{author}{\bibinfo{person}{Armin Alaghi} {and} \bibinfo{person}{John~P
  Hayes}.} \bibinfo{year}{2013}\natexlab{}.
\newblock \showarticletitle{Exploiting correlation in stochastic circuit
  design}. In \bibinfo{booktitle}{\emph{2013 IEEE 31st International Conference
  on Computer Design (ICCD)}}. IEEE, \bibinfo{pages}{39--46}.
\newblock


\bibitem[ANSYS(2003)]%
        {lumerical}
\bibfield{author}{\bibinfo{person}{ANSYS}.} \bibinfo{year}{2003}\natexlab{}.
\newblock \bibinfo{title}{Lumerical}.
\newblock
\newblock
\urldef\tempurl%
\url{http://www.lumerical.com/products}
\showURL{%
\tempurl}


\bibitem[Appeltant et~al\mbox{.}(2011)]%
        {Appeltant2011InformationPU}
\bibfield{author}{\bibinfo{person}{Lennert Appeltant} {et~al\mbox{.}}}
  \bibinfo{year}{2011}\natexlab{}.
\newblock \showarticletitle{Information processing using a single dynamical
  node as complex system}.
\newblock \bibinfo{journal}{\emph{Nature Communications}}  \bibinfo{volume}{2}
  (\bibinfo{year}{2011}).
\newblock


\bibitem[Bahadori et~al\mbox{.}(2017)]%
        {BahadoriDATE2017}
\bibfield{author}{\bibinfo{person}{Meisam Bahadori} {et~al\mbox{.}}}
  \bibinfo{year}{2017}\natexlab{}.
\newblock \showarticletitle{Energy-performance optimized design of silicon
  photonic interconnection networks for high-performance computing}. In
  \bibinfo{booktitle}{\emph{DATE}}.
\newblock


\bibitem[Bangari et~al\mbox{.}(2020)]%
        {deapcnn}
\bibfield{author}{\bibinfo{person}{Viraj Bangari} {et~al\mbox{.}}}
  \bibinfo{year}{2020}\natexlab{}.
\newblock \showarticletitle{Digital Electronics and Analog Photonics for
  Convolutional Neural Networks ({DEAP}-{CNNs})}.
\newblock \bibinfo{journal}{\emph{JSTQE}} (\bibinfo{year}{2020}).
\newblock


\bibitem[Bogaerts et~al\mbox{.}(2012)]%
        {MRR-tutorial}
\bibfield{author}{\bibinfo{person}{W. Bogaerts} {et~al\mbox{.}}}
  \bibinfo{year}{2012}\natexlab{}.
\newblock \showarticletitle{Silicon microring resonators}.
\newblock \bibinfo{journal}{\emph{Laser \& Photonics Reviews}}
  \bibinfo{volume}{6}, \bibinfo{number}{1} (\bibinfo{year}{2012}),
  \bibinfo{pages}{47--73}.
\newblock


\bibitem[Brückerhoff-Plückelmann et~al\mbox{.}(2022)]%
        {BhaskaranPCA2022}
\bibfield{author}{\bibinfo{person}{Frank Brückerhoff-Plückelmann}
  {et~al\mbox{.}}} \bibinfo{year}{2022}\natexlab{}.
\newblock \showarticletitle{A large scale photonic matrix processor enabled by
  charge accumulation}.
\newblock \bibinfo{journal}{\emph{Nanophotonics}} (\bibinfo{year}{2022}).
\newblock


\bibitem[Chen et~al\mbox{.}(2021)]%
        {Chen_2021_ICCV}
\bibfield{author}{\bibinfo{person}{Weihan Chen} {et~al\mbox{.}}}
  \bibinfo{year}{2021}\natexlab{}.
\newblock \showarticletitle{Towards Mixed-Precision Quantization of Neural
  Networks via Constrained Optimization}. In \bibinfo{booktitle}{\emph{ICCV}}.
\newblock


\bibitem[Duport et~al\mbox{.}(2016)]%
        {opticalmziDFRC}
\bibfield{author}{\bibinfo{person}{François Duport} {et~al\mbox{.}}}
  \bibinfo{year}{2016}\natexlab{}.
\newblock \showarticletitle{Fully analogue photonic reservoir computer}.
\newblock \bibinfo{journal}{\emph{Scientific Reports}} (\bibinfo{year}{2016}).
\newblock


\bibitem[Jaeger(2002)]%
        {narma10}
\bibfield{author}{\bibinfo{person}{Herbert Jaeger}.}
  \bibinfo{year}{2002}\natexlab{}.
\newblock \showarticletitle{Adaptive Nonlinear System Identification with Echo
  State Networks}. In \bibinfo{booktitle}{\emph{NIPS}}.
\newblock


\bibitem[Jaeger et~al\mbox{.}(2004)]%
        {channelequalization}
\bibfield{author}{\bibinfo{person}{Herbert Jaeger} {et~al\mbox{.}}}
  \bibinfo{year}{2004}\natexlab{}.
\newblock \showarticletitle{Harnessing Nonlinearity: Predicting Chaotic Systems
  and Saving Energy in Wireless Communication}.
\newblock \bibinfo{journal}{\emph{Science}} (\bibinfo{year}{2004}).
\newblock


\bibitem[Karempudi et~al\mbox{.}(2022)]%
        {praneeth4pam}
\bibfield{author}{\bibinfo{person}{Venkata Sai~Praneeth Karempudi}
  {et~al\mbox{.}}} \bibinfo{year}{2022}\natexlab{}.
\newblock \showarticletitle{Photonic Networks-on-Chip Employing Multilevel
  Signaling: A Cross-Layer Comparative Study}.
\newblock \bibinfo{journal}{\emph{JETCS}} (\bibinfo{year}{2022}).
\newblock


\bibitem[Karempudi et~al\mbox{.}(2021)]%
        {karempudi2021}
\bibfield{author}{\bibinfo{person}{Venkata Sai~Praneeth Karempudi},
  \bibinfo{person}{Shreyan Datta}, {and} \bibinfo{person}{Ishan~G Thakkar}.}
  \bibinfo{year}{2021}\natexlab{}.
\newblock \showarticletitle{Design Exploration and Scalability Analysis of a
  CMOS-Integrated, Polymorphic, Nanophotonic Arithmetic-Logic Unit}. In
  \bibinfo{booktitle}{\emph{Proceedings of the 19th ACM Conference on Embedded
  Networked Sensor Systems}}. \bibinfo{pages}{628--634}.
\newblock


\bibitem[{Karempudi \em et al.}(2023)]%
        {karempudi2023}
\bibfield{author}{\bibinfo{person}{V. {Karempudi \em et al.}}}
  \bibinfo{year}{2023}\natexlab{}.
\newblock \showarticletitle{A Polymorphic Electro-Optic Logic Gate for
  High-Speed Reconfigurable Computing Circuits}.
\newblock \bibinfo{journal}{\emph{arXiv preprint arXiv:2301.13626}}
  (\bibinfo{year}{2023}).
\newblock


\bibitem[Liu et~al\mbox{.}(2019)]%
        {holylight}
\bibfield{author}{\bibinfo{person}{Weichen Liu} {et~al\mbox{.}}}
  \bibinfo{year}{2019}\natexlab{}.
\newblock \showarticletitle{HolyLight: A Nanophotonic Accelerator for Deep
  Learning in Data Centers}. In \bibinfo{booktitle}{\emph{DATE}}.
\newblock


\bibitem[Luan et~al\mbox{.}(2023)]%
        {Luan2023Nature}
\bibfield{author}{\bibinfo{person}{Enxiao Luan} {et~al\mbox{.}}}
  \bibinfo{year}{2023}\natexlab{}.
\newblock \showarticletitle{Towards a high-density photonic tensor core enabled
  by intensity-modulated microrings and photonic wire bonding}.
\newblock \bibinfo{journal}{\emph{Scientific Reports}} (\bibinfo{year}{2023}).
\newblock


\bibitem[{Qiu \em et al.}(2012)]%
        {qiu2012}
\bibfield{author}{\bibinfo{person}{C. {Qiu \em et al.}}}
  \bibinfo{year}{2012}\natexlab{}.
\newblock \showarticletitle{Demonstration of reconfigurable electro-optical
  logic with silicon photonic integrated circuits}.
\newblock \bibinfo{journal}{\emph{Optics letters}} \bibinfo{volume}{37},
  \bibinfo{number}{19} (\bibinfo{year}{2012}), \bibinfo{pages}{3942--3944}.
\newblock


\bibitem[Shastri et~al\mbox{.}(2021)]%
        {SiPh-NN-AI-Survey}
\bibfield{author}{\bibinfo{person}{B.J. Shastri} {et~al\mbox{.}}}
  \bibinfo{year}{2021}\natexlab{}.
\newblock \showarticletitle{Photonics for artificial intelligence and
  neuromorphic computing}.
\newblock \bibinfo{journal}{\emph{Nat. Photonics}}  \bibinfo{volume}{15}
  (\bibinfo{year}{2021}), \bibinfo{pages}{102–114}.
\newblock


\bibitem[Shiflett et~al\mbox{.}(2020)]%
        {pixel}
\bibfield{author}{\bibinfo{person}{Kyle Shiflett} {et~al\mbox{.}}}
  \bibinfo{year}{2020}\natexlab{}.
\newblock \showarticletitle{PIXEL: Photonic Neural Network Accelerator}. In
  \bibinfo{booktitle}{\emph{HPCA}}.
\newblock


\bibitem[{Shiflett\em et al.}(2021)]%
        {shiflett2021}
\bibfield{author}{\bibinfo{person}{K. {Shiflett\em et al.}}}
  \bibinfo{year}{2021}\natexlab{}.
\newblock \showarticletitle{Bitwise Neural Network Acceleration Using Silicon
  Photonics}. In \bibinfo{booktitle}{\emph{GLSVLSI}}. \bibinfo{pages}{9--14}.
\newblock


\bibitem[Shoman et~al\mbox{.}(2018)]%
        {ShomanOFC2018}
\bibfield{author}{\bibinfo{person}{Hossam Shoman} {et~al\mbox{.}}}
  \bibinfo{year}{2018}\natexlab{}.
\newblock \showarticletitle{Compact Silicon Microring Modulator with Tunable
  Extinction Ratio and Wide FSR}. In \bibinfo{booktitle}{\emph{OFC}}.
\newblock


\bibitem[Sludds et~al\mbox{.}(2022)]%
        {alexandermit2022}
\bibfield{author}{\bibinfo{person}{Alexander Sludds} {et~al\mbox{.}}}
  \bibinfo{year}{2022}\natexlab{}.
\newblock \showarticletitle{Delocalized photonic deep learning on the
  internet's edge}.
\newblock \bibinfo{journal}{\emph{Science}} (\bibinfo{year}{2022}).
\newblock


\bibitem[Solow(1994)]%
        {santfe}
\bibfield{author}{\bibinfo{person}{Andrew~R. Solow}.}
  \bibinfo{year}{1994}\natexlab{}.
\newblock \showarticletitle{Forecasting the Future and Understanding the Past}.
\newblock \bibinfo{journal}{\emph{Science}} (\bibinfo{year}{1994}).
\newblock


\bibitem[Sri~Vatsavai and Thakkar(2023)]%
        {sri2023bit}
\bibfield{author}{\bibinfo{person}{Sairam Sri~Vatsavai} {and}
  \bibinfo{person}{Ishan Thakkar}.} \bibinfo{year}{2023}\natexlab{}.
\newblock \showarticletitle{A Bit-Parallel Deterministic Stochastic
  Multiplier}.
\newblock \bibinfo{journal}{\emph{arXiv e-prints}} (\bibinfo{year}{2023}),
  \bibinfo{pages}{arXiv--2302}.
\newblock


\bibitem[Sri~Vatsavai and Thakkar(2022)]%
        {cases2022}
\bibfield{author}{\bibinfo{person}{Sairam Sri~Vatsavai} {and}
  \bibinfo{person}{Ishan~G. Thakkar}.} \bibinfo{year}{2022}\natexlab{}.
\newblock \showarticletitle{Photonic Reconfigurable Accelerators for Efficient
  Inference of CNNs With Mixed-Sized Tensors}.
\newblock \bibinfo{journal}{\emph{TCAD}} (\bibinfo{year}{2022}).
\newblock


\bibitem[Sunny et~al\mbox{.}(2021)]%
        {robin}
\bibfield{author}{\bibinfo{person}{Febin~P. Sunny} {et~al\mbox{.}}}
  \bibinfo{year}{2021}\natexlab{}.
\newblock \showarticletitle{ROBIN: A Robust Optical Binary Neural Network
  Accelerator}.
\newblock \bibinfo{journal}{\emph{ACM Trans. Embed. Comput. Syst.}}
  (\bibinfo{year}{2021}).
\newblock


\bibitem[Vatsavai et~al\mbox{.}(2020)]%
        {proteusnocs2020}
\bibfield{author}{\bibinfo{person}{Sairam~Sri Vatsavai} {et~al\mbox{.}}}
  \bibinfo{year}{2020}\natexlab{}.
\newblock \showarticletitle{PROTEUS: Rule-Based Self-Adaptation in Photonic
  NoCs for Loss-Aware Co-Management of Laser Power and Performance}. In
  \bibinfo{booktitle}{\emph{NOCS}}.
\newblock


\bibitem[Vatsavai et~al\mbox{.}(2023a)]%
        {oxbnn}
\bibfield{author}{\bibinfo{person}{Sairam~Sri Vatsavai} {et~al\mbox{.}}}
  \bibinfo{year}{2023}\natexlab{a}.
\newblock \bibinfo{title}{An Optical XNOR-Bitcount Based Accelerator for
  Efficient Inference of Binary Neural Networks}.
\newblock
\newblock
\showeprint[arxiv]{2302.06405}~[cs.AR]


\bibitem[Vatsavai et~al\mbox{.}(2023b)]%
        {sconna}
\bibfield{author}{\bibinfo{person}{Sairam~Sri Vatsavai} {et~al\mbox{.}}}
  \bibinfo{year}{2023}\natexlab{b}.
\newblock \bibinfo{title}{SCONNA: A Stochastic Computing Based Optical
  Accelerator for Ultra-Fast, Energy-Efficient Inference of Integer-Quantized
  CNNs}.
\newblock
\newblock
\showeprint[arxiv]{2302.07036}~[cs.AR]


\bibitem[Vatsavai and Thakkar(2021)]%
        {SairamVLSID2021}
\bibfield{author}{\bibinfo{person}{Sairam~Sri Vatsavai} {and}
  \bibinfo{person}{Ishan Thakkar}.} \bibinfo{year}{2021}\natexlab{}.
\newblock \showarticletitle{Silicon Photonic Microring Based Chip-Scale
  Accelerator for Delayed Feedback Reservoir Computing}. In
  \bibinfo{booktitle}{\emph{VLSID}}.
\newblock


\bibitem[Ying et~al\mbox{.}(2020)]%
        {ying2020}
\bibfield{author}{\bibinfo{person}{Zhoufeng Ying}, \bibinfo{person}{Chenghao
  Feng}, \bibinfo{person}{Zheng Zhao}, \bibinfo{person}{Shounak Dhar},
  \bibinfo{person}{Hamed Dalir}, \bibinfo{person}{Jiaqi Gu},
  \bibinfo{person}{Yue Cheng}, \bibinfo{person}{Richard Soref},
  \bibinfo{person}{David~Z Pan}, {and} \bibinfo{person}{Ray~T Chen}.}
  \bibinfo{year}{2020}\natexlab{}.
\newblock \showarticletitle{Electronic-photonic arithmetic logic unit for
  high-speed computing}.
\newblock \bibinfo{journal}{\emph{Nature communications}} \bibinfo{volume}{11},
  \bibinfo{number}{1} (\bibinfo{year}{2020}), \bibinfo{pages}{2154}.
\newblock


\bibitem[{Ying \em et al.}(2019)]%
        {ying2019integrated}
\bibfield{author}{\bibinfo{person}{Z {Ying \em et al.}}}
  \bibinfo{year}{2019}\natexlab{}.
\newblock \showarticletitle{Integrated multi-operand electro-optic logic gates
  for optical computing}.
\newblock \bibinfo{journal}{\emph{APL}} (\bibinfo{year}{2019}).
\newblock


\bibitem[{Zoakee \em et al.}(2020)]%
        {lightbulb}
\bibfield{author}{\bibinfo{person}{F. {Zoakee \em et al.}}}
  \bibinfo{year}{2020}\natexlab{}.
\newblock \showarticletitle{LightBulb: A photonic-nonvolatile-memory-based
  accelerator for binarized convolutional neural networks}. In
  \bibinfo{booktitle}{\emph{2020 DATE}}. IEEE, \bibinfo{pages}{1438--1443}.
\newblock


\end{thebibliography}

\end{document}